\documentclass[twocolumn,showpacs,preprintnumbers,amsmath,amssymb]{revtex4}
\usepackage{graphicx}
\usepackage{dcolumn}
\usepackage{bm}
\usepackage{longtable}
\def\bea {\begin{eqnarray}}
\def\eea {\end{eqnarray}}

\def\be {\begin{equation}}
\def\ee {\end{equation}}

\begin{document}
\title{Band structures and intruder $\pi$$i_{13/2}$ state in $^{197}$Tl}
\author{H. Pai$^{1}$}
\altaffiliation{Present Address: Institut f\"ur Kernphysik, Technische Universit\"at Darmstadt, GERMANY}
\email{hari.vecc@gmail.com, hpai@ikp.tu-darmstadt.de}
\author{G. Mukherjee$^{1}$}
\thanks{Corresponding author}
\email{gopal@vecc.gov.in}
\author{S. Bhattacharya$^{1}$}
\author{C. Bhattacharya$^1$}
\author{S. Bhattacharyya$^1$}
\author{T. Bhattacharjee$^1$}
\author{S. Chanda$^2$}
\author{S. Rajbanshi$^{3}$}
\author{A. Goswami$^{3}$}
\author{M.R. Gohil$^{1}$}
\author{S. Kundu$^{1}$}
\author{T.K. Ghosh$^{1}$}
\author{K. Banerjee$^{1}$}
\author{T.K. Rana$^{1}$}
\author{R. Pandey$^{1}$}
\author{G.K. Prajapati$^{1}$}
\altaffiliation{Present Address: Nuclear Physics Division, Bhabha Atomic Research Centre, Mumbai - 400085, INDIA}
\author{S. R. Banerjee$^{1}$}
\author{S. Mukhopadhyay$^{1}$}
\author{D. Pandit$^{1}$}
\author{S. Pal$^{1}$}
\author{J. K. Meena$^{1}$}
\author{P. Mukhopadhyay$^{1}$}
\author{A. Chawdhury$^{1}$}
\affiliation{%
$^1$Variable Energy Cyclotron Centre, 1/AF Bidhan Nagar, Kolkata 700064, INDIA\\
$^2$Physics Department, Fakir Chand College, Diamond Harbour, West Bengal, INDIA\\
$^3$Saha Institute of Nuclear Physics, Kolkata 700064, INDIA}
\author{}
\affiliation{
\\
 forced
}%

\date{\today}
\begin{abstract}
The excited states in the odd-$A$ $^{197}$Tl nucleus have been studied by populating them using the 
$^{197}$Au($\alpha$, 4$n$)$^{197}$Tl reaction at the beam energy of 48 MeV. The $\gamma-\gamma$ coincidence data
were taken using a combination of clover, LEPS and single crystal HPGe detectors. Precise spin and parity 
assignments of the excited states have been done through the polarization and the DCO measurements. A new 
band structure has been identified and the evidence for a possible intruder $\pi i_{13/2}$ state has been 
found for the first time. Possible configurations of the observed bands have been discussed. The total 
Routhian surface calculations have been performed to study the shape of $^{197}$Tl for different configurations.

\end{abstract}
\pacs{21.10.Re; 21.10.-k; 23.20.Lv; 23.20.En; 21.60.Cs; 27.80.+w }

\maketitle
\section{Introduction}
The ground state of the odd-$A$ Tl ($Z=81$) nuclei are 1/2$^+$~\cite{aodd-Tl1,aodd-Tl2,aodd-Tl3} 
corresponding to the proton hole in the 3s$_{1/2}$ (below the Z = 82 spherical shell closure) orbital. 
The low-lying excited states in the odd-$A$ thallium nuclei in the $A = 190$ region have been interpreted 
by the occupation of the odd proton in the $\pi$$d_{3/2}$ and $\pi$$d_{5/2}$ orbitals. However, the 
``intruder" $\pi$$h_{9/2}$ and $\pi$$i_{13/2}$ orbitals are required to describe the higher spin 
levels in these nuclei. These orbitals intrude from the major shell above $Z = 82$ in to the shell 
below it for both prolate and oblate deformations. These orbitals play significant role in breaking 
the spherical symmetry in nuclei by inducing non-spherical shapes in them. It is important to note 
that the $\pi i_{13/2}$ level lies above the $Z = 92$ spherical sub-shell closure. Therefore, the 
``intruder" $\pi i_{13/2}$ level in the lighter Tl nuclei provides a playground to study the properties 
of the levels for the heavy nuclei above $Z = 92$ which are otherwise difficult to study.

Rotational bands based on the intruder $\pi$[505]9/2$^-$ Nilsson state with oblate deformation and 
decoupled band based on the $\pi h_{9/2}$ intruder level with prolate deformation have been reported in 
the odd-$A$ $^{189-197}$Tl nuclei~\cite{a1,a2,a3,a4,new74}. Similarly, the strongly coupled oblate band based 
on the $\pi$[606]13/2$^+$ Nilsson state and weakly coupled prolate band from the $\pi$$i_{13/2}$ orbital 
have also been observed in the neutron deficient (N $<$ 114) isotopes $^{191,193}$Tl~\cite{a4,new74,a5}. 
However, the excited state corresponding to the intruder $\pi$$i_{13/2}$ orbital has not yet been observed in 
$^{195,197}$Tl. It has been seen that the excitation energy of the 13/2$^+$ state corresponding to 
the $\pi i_{13/2}$ orbital in the odd-$A$ thallium nuclei increases gradually with the neutron 
number~\cite{TOIA} and hence, it may become non-yrast for the heavy thallium isotopes. The non-yrast 
states can be better studied using light-ion beams, like $\alpha$-induced fusion-evaporation reactions. 
In the present work, the $\gamma$-ray spectroscopy of $^{197}$Tl has been studied using $\alpha$ beam 
in order to investigate the proton intruder states along with the multi-quasiparticle states which may 
be originated from the coupling of the odd-proton with the aligned neutron pairs in this nucleus. To 
identify the intruder levels, it is necessary to deduce the spin and parities of the excited states 
unambiguously. So, a clover HPGe detector was used as a $\gamma$-ray polarimeter to deduce the multipolarity 
of the $\gamma$ rays which helps in the unambiguous determination of the parity of the excited states.

The high spin level structures in $^{197}$Tl were earlier investigated by R. M. Lieder et al.~\cite{a1} 
way back in 1978. They had used two Ge(Li) detectors to detect the $\gamma$ rays in coincidence and for 
angular distribution study. They have proposed a level scheme from these studies which included a band 
based on $\pi h_{9/2}$ orbital and a three-quasiparticle band.

\section{Experimental Method and Data Analysis}
In the present work, the excited states in $^{197}$Tl were populated by fusion-evaporation reaction 
$^{197}$Au($^{4}$He, 4$n$)$^{197}$Tl using $\alpha$ beam of energy 48 MeV from the K-130 cyclotron at 
Variable Energy Cyclotron Centre, Kolkata. A 5 mg/cm$^2$ self supporting $^{197}$Au target was used 
in this experiment. The experimental set up consisted of a single-crystal large high purity Ge (HPGe) 
detector (80\% relative efficiency), a clover HPGe detector and a LEPS detector which were placed at 
30$^\circ$, 90$^\circ$ and 135$^\circ$ angles, respectively, with respect to the beam direction. The 
detectors were arranged in a median plane configuration. A 50-element (25 each on the top and on the 
bottom) $BaF_{2}$ multiplicity array was also used. The energy and the timing information from each 
of the Ge detector and the multiplicity fold information were recorded in the list mode using a 
$\gamma$-$\gamma$ trigger. There was no hardware condition set on multiplicity fold. A hardware 
Time-to-Amplitude Converter (TAC) module was used to record the time between the master trigger and the 
RF signal of the cyclotron (RF-$\gamma$ TAC) to identify the ``beam-on" and ``beam-off" events. About 
4.3 $\times$ 10$^7$ coincidence events were recorded in this experiment. The detectors were calibrated 
for $\gamma$-ray energies and efficiencies by using the $^{133}$Ba and $^{152}$Eu radioactive sources. 
The coincidence data were sorted into a $\gamma$-$\gamma$ matrix using the data from the single-crystal 
HPGe and clover HPGe detectors for offline analysis. This two-dimensional matrix was created with a gate 
on the prompt peak in the RF-$\gamma$ TAC. A prompt time window of $\pm$50 ns was chosen. The analysis has 
been done by using the program RADWARE~\cite{a7}.

The multipolarities of the $\gamma$-ray transitions have been determined from the angular correlation 
analysis using the method of directional correlation from the oriented states (DCO) ratio, following 
the prescriptions of Kr\"{a}mer-Flecken et al.~\cite{bkm}. For the DCO ratio analysis, the coincidence 
events were sorted into an asymmetry matrix with data from 30$^\circ$ detector ($\theta_2$) on one axis 
and 90$^\circ$ detector ($\theta_1$) on the other axis. 
The DCO ratio (for $\gamma_1$, gated by a $\gamma$-ray $\gamma_2$ of known multipolarity) is obtained 
from the intensities of the $\gamma$ rays (I$_\gamma$) at two angles $\theta_1$ and $\theta_2$, as
\begin{equation}\label{rdco}
R_{DCO} = \frac{I_{\gamma_1} ~ at~ \theta_1, ~gated ~by ~\gamma_2 ~at ~\theta_2}
               {I_{\gamma_1} ~at ~\theta_2 ~gated ~by ~\gamma_2 ~at ~\theta_1}
\end{equation}
By putting gates on the transitions with known multipolarity along the two axes of the above matrix, the 
DCO ratios were obtained for each $\gamma$ ray. For stretched transitions, the value of R$_{DCO}$ would be 
close to unity for the same multipolarity of $\gamma_1$ and $\gamma_2$. For different multipolarities and 
for mixed transitions, the value of R$_{DCO}$ depends on the detector angles ($\theta_1$ and $\theta_2$) 
and on the mixing ratio ($\delta$). 

The R$_{DCO}$ values were also calculated using the `angcor' code \cite{angcor} and were compared with the
measured ones. The spin alignment parameter  $\sigma /J$ is important for such calculations which is normally
taken as 0.3 for heavy-ion fusion-evaporation reaction. However, for the light-ion induced reaction, as in the
present case, the $\sigma /J$ value is expected to differ. Therefore, the value of the $\sigma /J$ parameter
has been obtained in the present work by comparing the measured value with the calculated values of R$_{DCO}$ 
of known multipole transitions in $^{197}$Tl. For this, the 685-keV stretched E2 transition was chosen gated by 
the 560.9-keV E1 transition. This choice was based on the fact that both are pure transitions of different 
multipolarities. The gate width was taken sufficiently narrow so that there is no contamination. The R$_{DCO}$ 
value for this has been calculated for various values of $\sigma /J$ and are shown in Fig.~1. It can be seen 
from this figure that a value of $\sigma /J = 0.39$ reproduces the mean experimental value of $R_{DCO} = 0.61$. 
This value of $\sigma /J$ was used to calculate the value of R$_{DCO}$ for the other transitions. For a pure 
dipole transition, gated by a stretched quadrupole transition, the calculated value of R$_{DCO}$ is 1.64 while 
for a mixed dipole transition with $\delta = 1.0$, R$_{DCO}$ is 0.71. 

The validity of the R$_{DCO}$ measurements was further tested by comparing the experimental values, obtained 
in this work for the in-band $\Delta I=1$ mixed $M1+E2$ transitions, for which the mixing ratios are known from 
Lieder et al. \cite{a1}, gated by the stretched $E2$ transitions in $^{197}$Tl with the calculated ones. These 
agree well within the uncertainty of our measurements and the uncertainty of the reported mixing ratios.

\begin{figure}[!]
\begin{center}
\includegraphics*[scale=0.3, angle = 0]{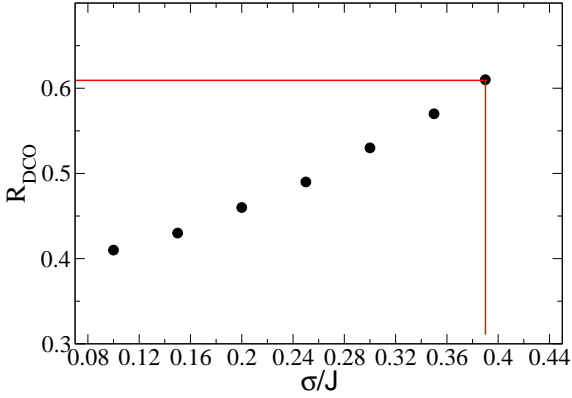}
\caption{(Color online) Calculated values of the R$_{DCO}$ for different values of 
$\sigma /J$ for the 685 keV $E2$ transition in $^{197}$Tl gated by the 561 keV $E1$ transition.
The solid ines correspond to the $\sigma /J$ value which reproduces the measured R$_{DCO}$ values
for the above transitions.} 
\label{Fig1}
\end{center}
\end{figure}

Definite parities of the excited states have been assigned from the polarization asymmetry (PDCO) ratio, 
as described in Refs.~\cite{bp1,bp2,rp}, from the parallel and perpendicular scattering of a $\gamma$-ray 
photon inside the clover detector, used in the present experiment. 
The PDCO ratio measurement gives a qualitative idea about the type of a transition ($E/M$). The PDCO 
asymmetry parameters have been deduced using the relation,
\begin{equation}\label{pdco}
\Delta_{PDCO} = \frac{a(E_\gamma) N_\perp - N_\parallel}{a(E_\gamma)N_\perp + N_\parallel},
\end{equation}
where $N_\parallel$ and $N_\perp$ are the counts for the actual Compton scattered $\gamma$ rays in the 
planes parallel and perpendicular to the reaction plane. 
The parallel and perpendicular scattering data of the $\gamma$ rays were obtained from the clover detector
which was placed at 90$^o$ for better polarization sensitivity. For the parallel counts (N$_\parallel$), the
condition is set on the selection of events in which the  $\gamma$ rays recorded by a clover detector 
undergoes scattering between the segments parallel to the beam axis (orientation axis). On the other hand, 
for perpendicular counts (N$_\perp$), the events scatter between the segments perpendicular to the beam 
axis were taken. The correction due to possible asymmetry and different response of the clover segments, 
a correction factor, defined by a($E_\gamma$) = $\frac{N_\parallel} {N_\perp}$, was determined using
an unpolarized $^{152}$Eu source. The values of a($E_\gamma$) have been fitted using the expression
a($E_\gamma$) = a + b$E_\gamma$. The fitting, shown in Fig.~2,  gives the values of the constants as 
a = 0.990(12) and b $\sim$ 10$^{-8}$.

\begin{figure}[!]
\begin{center}
\includegraphics*[scale=0.3, angle = 0]{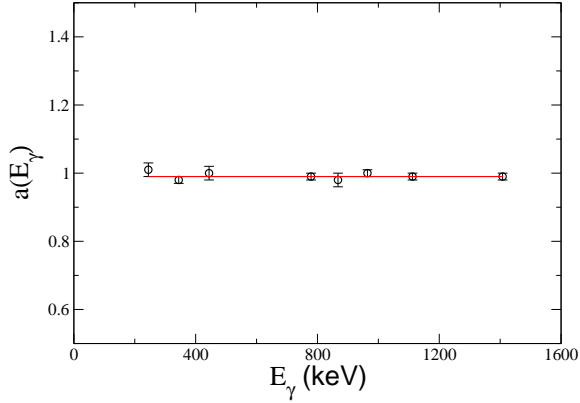}
\caption{(Color online) The asymmetry correction factor $a(E_\gamma$) at different $\gamma$ energies 
from $^{152}$Eu source. The solid line corresponds to a linear fit of the data. The fitted
value of the correction factor a($E_\gamma$) has been shown.}
\label{Fig2}
\end{center}
\end{figure}

Positive and negative values of $\Delta_{PDCO}$ indicate electric and magnetic type of transitions, 
respectively. The low energy cut off for the polarization measurement was about 200 keV in the present 
experiment. The asymmetry parameters ($\Delta_{PDCO}$) obtained, in the present experiment, for different 
$\gamma$-transitions are shown in Fig.~3. 

\begin{figure}[!]
\begin{center}
\includegraphics*[scale=0.3, angle = 0]{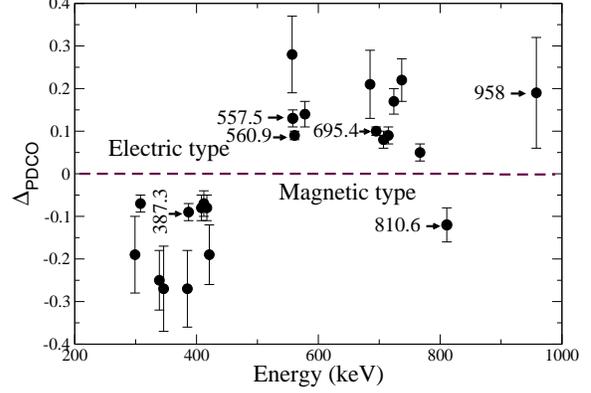}
\caption{Polarization asymmetry ($\Delta_{PDCO}$) for different transitions.}
\label{Fig3}
\end{center}
\end{figure}

\subsection{Experimental Results}
Single gated $\gamma$-ray spectra obtained in this work, gated by the previously known 387-keV and 
561-keV $\gamma$ rays belonging to $^{197}$Tl, are shown in Fig. 4(a) and Fig. 4(b), respectively. These 
spectra show all the known $\gamma$ lines reported by R. M. Lieder et al.~\cite{a1} and the new $\gamma$ 
rays observed in the present work. Tentatively assigned 470-keV transition from the 3067-keV level and 
535-keV transition from the 2801-keV level, reported in Ref.~\cite{a1}, have been clearly observed in 
Figs. 4(a) and 4(b), respectively. Therefore, these $\gamma$-ray transitions have been firmly placed in 
the level scheme. On the other hand, the 412.5-keV $\gamma$-ray, placed above the 3277-keV 29/2$^{(-)}$ 
state in Ref.\cite{a1}, has been excluded in the present work based on its coincidence relation.

\begin{figure*}[!]
\begin{center}
\includegraphics*[width=15cm,keepaspectratio=true, angle = 0]{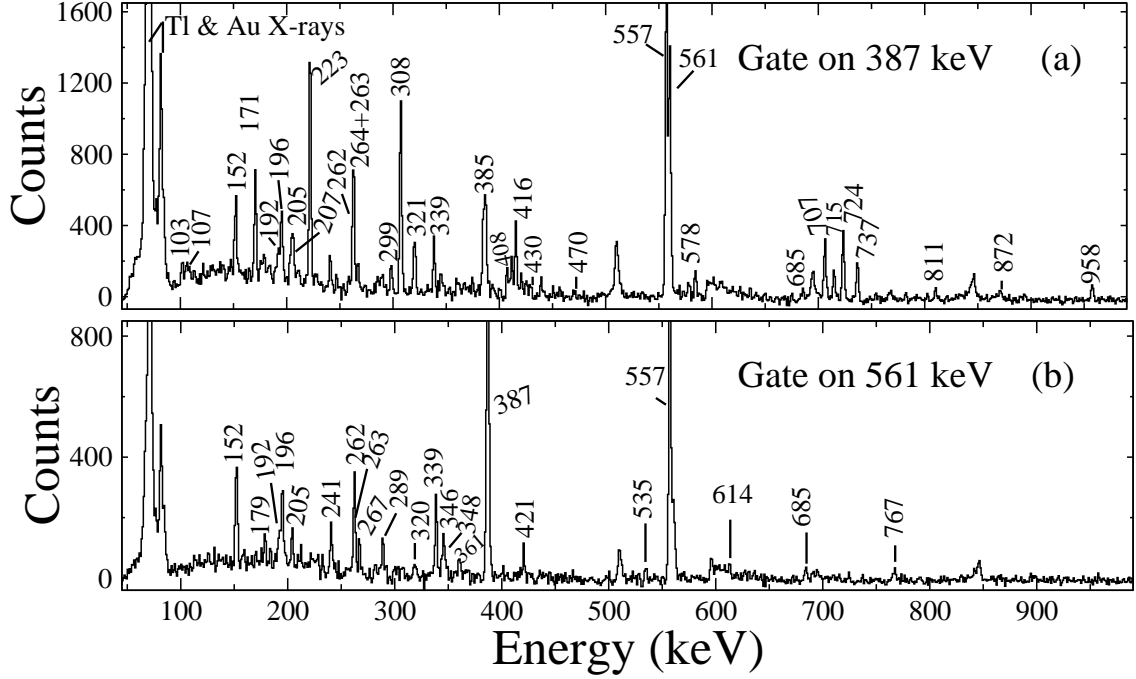}
\caption{Coincidence spectra gated by (a) 387-keV and (b) 561-keV $\gamma$ transitions in $^{197}$Tl.}
\label{Fig4}
\end{center}
\end{figure*}

%

An improved level scheme of $^{197}$Tl, obtained in this work, is given in Fig.~5. The ground state of 
$^{197}$Tl is known to be a long-lived (T$_{1/2}$ = 2.84 hour) 1/2$^{+}$ state \cite{ensdf}. An isomeric 
state ($J^\pi = 9/2^-$) with a halflife of $0.53 s$ is also known in this nucleus at 608 keV of excitation 
energy \cite{ensdf}. The deduced excitation energy, spin and parity of the excited states above the isomer 
and the multipolarity of the $\gamma$ rays, together with the other relevant information concerning 
their placement in the proposed level scheme, are summarized in Table~I.

\begin{longtable*}{ccccccc}
\caption{\label{tab:Table1}List of $\gamma$-rays belonging to $^{197}$Tl, as observed in the present work.
Energy (E$_\gamma$) and intensity (I$_\gamma$) of the $\gamma$-rays, energy (E$_i$), spin and parity (J$^\pi$)
of the level from which the $\gamma$-ray is emitted, the deduced DCO (R$_{DCO}$) and PDCO ($\Delta_{PDCO}$)
ratios and the adopted multipolarity of the $\gamma$-rays are tabulated. }\\
\hline
$E_{\gamma}$~ &~ $E_{i}$~ &~ $J^{\pi}_i$ & $I_{\gamma}$
$^{\footnotemark[1]}$~ &~ $R_{DCO}$~ &~$\Delta_{PDCO}$~ &~Multi- \\
(keV) & (keV) & & & (Err) & (Err) & polarity \\
\hline
\endfirsthead
\multicolumn{7}{c}{Table~{\ref{int-197Tl}}: Continued...}\\
\hline
$E_{\gamma}$~ &~ $E_{i}$~ &~ $J^{\pi}_i$ & $I_{\gamma}$
$^{\footnotemark[1]}$~ &~ $R_{DCO}$~ &~$\Delta_{PDCO}$~ &~Multi- \\
(keV) & (keV) & & & (Err) & (Err) & polarity \\
\hline
\endhead
\endfoot
\endlastfoot

 102.6(1)&3169.5  & $ 27/2^{-} $ & 2.4(3) & 1.90(34)$^{\footnotemark[2]}$  &    -    & M1+E2 \\

 107.2(3) & 3276.7 & $29/2^{-} $  &1.1(2) &1.41(34)$^{\footnotemark[2]}$ &   -     &   M1+E2 \\

 152.4(1)& 2266.1 & $17/2^{-} $ &6.8(7)& 1.67(34)$^{\footnotemark[3]}$&   -     & M1(+E2) \\

 171.1(1)& 2596.6 & $21/2^{-} $ &6.9(6)& 1.51(17)$^{\footnotemark[4]}$&   -     & M1+E2    \\

 179.3(1)& 2555.0 & $19/2^{(-)} $ &4.0(4)& 1.00(21)$^{\footnotemark[5]}$&   -     & (M1+E2)  \\

 192.2(2) & 3760.1 & $27/2^{-}$ &2.0(2) & 1.16(18)$^{\footnotemark[5]}$& -     &M1+E2 \\

 195.9(1)& 2461.7 & $19/2^{-} $ &5.1(3)& 1.08(12)$^{\footnotemark[5]}$&- & M1+E2 \\

 204.6(3)& 3964.7 & $29/2^{-} $ &1.8(2) & 1.12(16)$^{\footnotemark[5]}$&-& M1+E2\\

 206.6(2) & 3066.9& $25/2^{-}$ &6.2(7)& 1.33(18)$^{\footnotemark[4]}$  &- & M1+E2  \\

 222.6(1) & 608.1& $9/2^{-} $ &29.0(20)&- &-  &        E3$^{\footnotemark[6]}$ \\

 241.2(1)& 2796.1 & $21/2^{(-)} $ &6.8(9)& 1.03(15)$^{\footnotemark[5]}$&   -     & (M1+E2)  \\

 261.9(1) & 2375.7 & $17/2^{(-)} $ &7.3(9)& 1.15(16)$^{\footnotemark[7]}$ &-& (M1+E2) \\

 263.2(1) & 2529.3 & $19/2^{-} $ &5.1(6)& 0.89(8)$^{\footnotemark[8]}$ &-& M1+E2 \\

 263.8(1) & 2860.3 & $23/2^{-} $ &8.4(8)& 1.51(19)$^{\footnotemark[9]} $& -0.05(2)&M1+E2\\

 267.4(1) & 3063.6 & $23/2^{(-)} $ &6.5(6)& 1.01(6)$^{\footnotemark[5]}$ &-&(M1+E2)\\

 289.0(1) & 3352.6 & $25/2^{(-)} $ &6.5(6)& 0.99(15)$^{\footnotemark[5]}$ &-&(M1+E2)\\

 298.7(3) & 2018.1 & $17/2^{-} $ &2.3(2)& 1.52(25)$^{\footnotemark[4]}$ &-0.19(9)& M1+E2 \\

 307.8(2) &1303.2  & $13/2^{-}$ &18.2(15) &1.50(15)$^{\footnotemark[9]}$ &-0.07(2)    &M1+E2\\

 319.6(1) & 3672.2 & $27/2^{(-)} $ &6.2(6)& 1.00(12)$^{\footnotemark[5]}$ &-&(M1+E2)\\

 320.6(3) &  2040.0 & $17/2^{-}$ &4.0(3)& 1.34(18)$^{\footnotemark[4]}$   &   -0.13(5) &  M1+E2 \\

 339.1(2)& 2800.8 & $21/2^{-}$ &4.6(4)& 1.09(11)$^{\footnotemark[5]}$  &-0.25(7)& M1(+E2)\\

 345.8(2) &3146.7 & $23/2^{-} $ &1.8(2)& 1.15(11)$^{\footnotemark[5]}$&-0.27(10) & M1+E2 \\

 347.6(3) & 2461.7 & $19/2^{-} $ &0.4(1)&-   &-& E2\\

 360.8(1) & 4033.0 & $29/2^{(-)} $ &3.5(5)& 1.03(29)$^{\footnotemark[5]}$ &-&(M1+E2)\\

 385.2(1)&2425.5  & $19/2^{-} $ & 3.7(5)&1.78(17)$^{\footnotemark[10]}$ &-0.27(9)  &M1+E2 \\

 385.5(1) &385.5   & $3/2^{+} $&100(5)&  -                 &- & M1+E2$^{\footnotemark[6]}$ \\

 387.3(1) &995.4   & $11/2^{-} $ & 64.0(60)&1.39(12)$^{\footnotemark[9]}$ &-0.09(2) & M1+E2 \\
 
 407.7(2) &2425.5  & $19/2^{-} $ & 5.0(4)&1.56(15)$^{\footnotemark[9]}$ &-0.08(3)   &M1+E2 \\
     
 416.5(2) &1719.3  & $15/2^{-} $ & 31.0(40)&1.86(16)$^{\footnotemark[2]}$ &-0.08(3)  & M1+E2 \\

 421.2(1) &3567.9  & $25/2^{-} $ & 2.1(2)&1.08(15)$^{\footnotemark[5]}$ &-0.19(7)  & M1(+E2) \\

 429.7(1) &2383.1  & - & 0.44(5)&- &-  & - \\

 470.2(3) &3066.9  & $25/2^{-} $ & 0.5(1)&- &-  & E2 \\

 534.9(2)&2800.8 & $21/2^{-} $ & 0.5(1)&- &-  & E2 \\

 557.1(1)&2596.9  & $21/2^{-} $ & 3.6(5)&1.06(18)$^{\footnotemark[4]}$ &0.28(9)  & E2 \\

 557.5(2) &1552.9  & $13/2^{+} $& 39.0(30)&1.03(10)$^{\footnotemark[5]}$ &0.13(2)  & E1 \\

 560.9(2) &2113.8  & $15/2^{-} $ & 30.0(31)&1.76(35)$^{\footnotemark[3]}$ &0.09(1)  & E1 \\

 578.2(1) &2596.6  & $21/2^{-} $ & 6.5(6)&1.02(14)$^{\footnotemark[9]}$ &0.14(3)  & E2 \\

 613.6(3) &3760.1  & $27/2^{-} $ & 0.7(1)&    -   &        - &                E2 \\

 685.1(1) &3146.7  & $23/2^{-} $ & 0.5(1)&0.61(10)$^{\footnotemark[5]}$ &0.21(8)  & E2 \\

 695.4(1)&1303.2  & $13/2^{-} $ & 22.0(21)&1.03(8)$^{\footnotemark[9]}$ &0.10(1)  & E2 \\

 706.6(2) &2425.5  & $19/2^{-} $ & 7.4(8)&1.09(11)$^{\footnotemark[10]}$ &0.08(2)  & E2 \\

 714.8(2) &2018.1  & $17/2^{-} $ & 7.5(8)&1.00(9)$^{\footnotemark[10]}$ &0.09(2)  & E2 \\

 723.9(2) &1719.3  & $15/2^{-} $ & 7.7(8)&0.99(8)$^{\footnotemark[2]}$ &0.17(3)  & E2 \\

 737.2(2)&2040.0  & $17/2^{-} $ & 4.7(5)&1.01(10)$^{\footnotemark[10]}$ &0.22(5)  & E2 \\

 767.1(1)&3567.9  & $25/2^{-} $ & 1.6(2)&0.56(9)$^{\footnotemark[5]}$ &0.05(2)  & E2 \\

 810.6(2) &2113.8  & $15/2^{-} $& 4.0(4)&1.95(28)$^{\footnotemark[10]}$ &-0.12(4)  & M1+E2\\ 

 871.2(1) &1866.6  & $13/2^{-} $& 1.1(1)&1.43(45)$^{\footnotemark[11]}$ &-  & M1+E2\\

 958.1(1) &1953.4  & $13/2^{+} $& 3.1(1)&1.02(17)$^{\footnotemark[11]}$ &0.19(13)  & E1\\

\hline
\label{int-197Tl}
\footnotetext[1]{Relative $\gamma$ ray intensities are estimated from singles \\measurement and normalized to 100 for the total\\ intensity 
of 385.5 keV $\gamma$ rays.}
\footnotetext[2]{From 706.6 keV (E2) DCO gate;}
\footnotetext[3]{From 767.1 keV (E2) DCO gate;}
\footnotetext[4]{From 723.9 keV (E2) DCO gate;}
\footnotetext[5]{From 560.9 keV (E1) DCO gate;}
\footnotetext[6]{From ref.~\cite{a1};}
\footnotetext[7]{From 241.2 keV (M1+E2) DCO gate;}
\footnotetext[8]{From 152.4 keV (M1(+E2)) DCO gate;}
\footnotetext[9]{From 714.8 keV (E2) DCO gate;}
\footnotetext[10]{From 695.4 keV (E2) DCO gate;}
\footnotetext[11]{From 387.3 keV (M1+E2) DCO gate;}
\end{longtable*}
\normalsize

\begin{figure*}[!]
\begin{center}
\includegraphics*[width=9cm,keepaspectratio=true, angle = -90]{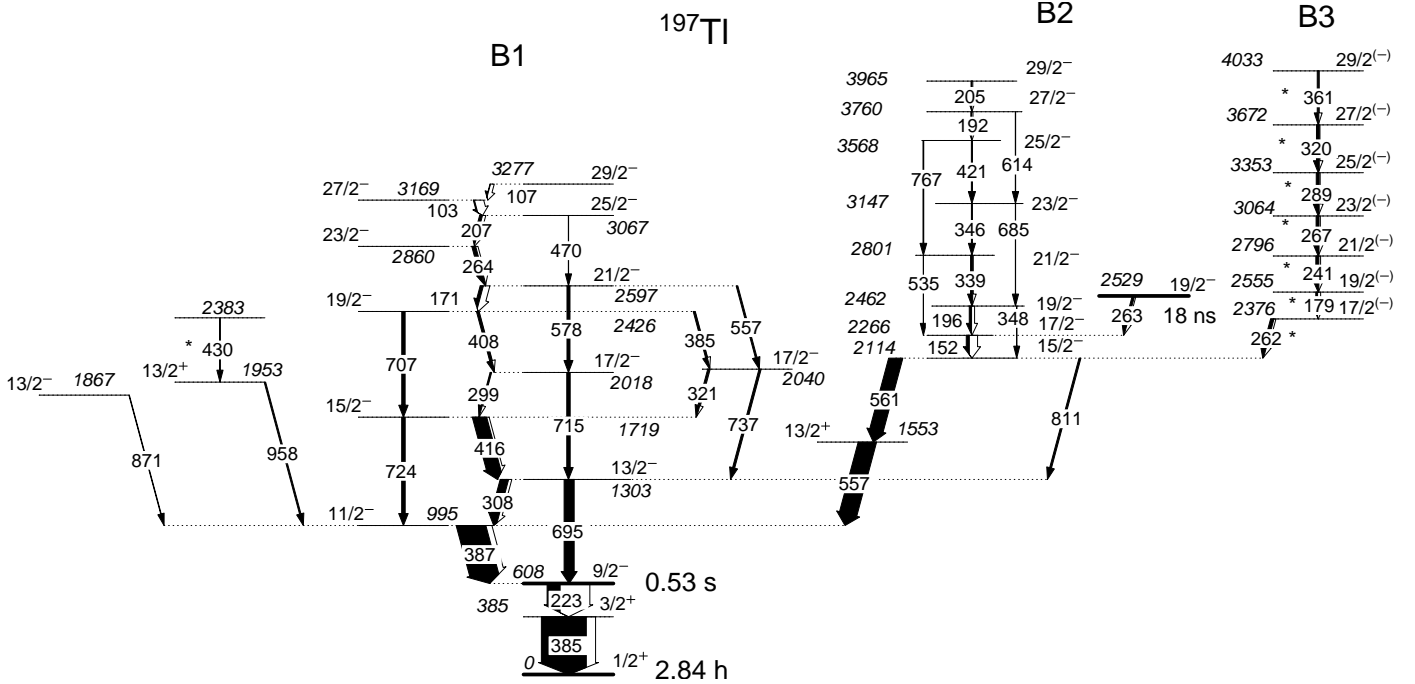}
\caption{Level scheme of $^{197}$Tl as proposed from the present work. The new $\gamma$-rays are
indicated by asterisks.}
\label{Fig5}
\end{center}
\end{figure*}

The relative intensity of the $\gamma$ rays were obtained from the singles spectrum of the clover 
detector with proper efficiency correction.  However, the relative intensities for the doublets at 
557.1-, 557.5-; 319.6, 320.6 and the triplets at 261.9-, 263.2-, 263.8-keV were separately obtained 
from gated spectra with proper normalization with the singles measurement. Gated spectra were used to 
obtain the relative intensity of the 385.2-keV transition as well.

The spin and the parity of the states were deduced in the previous work from the multipolarities of the 
$\gamma$ rays of $^{197}$Tl, determined from the angular distribution measurements~\cite{a1,a1a} and 
also from the measured conversion electron spectra \cite{a1}.

In the present work, the multipolarities of most of the known transitions in $^{197}$Tl have been well 
reproduced from the R$_{DCO}$ and the $\Delta_{PDCO}$ measurements. For example, the positive and the 
negative values of the $\Delta_{PDCO}$ ratios, respectively for the 695- and 387-keV $\gamma$ rays, 
(shown in Fig. 3) belonging to the lowest two transitions in the $9/2^-$ band indicate that they are 
predominantly electric and magnetic in character, respectively. The R$_{DCO}$ values for these two
transitions ($1.03\pm 0.08$ and $1.39 \pm 0.12$, respectively) gated by the 715-keV stretched $E2$ 
transition (see Table~I) indicate that these are predominantly quadrupole and dipole transitions, 
respectively. The $E2$ and $M1+E2$ character reported in Ref.\cite{a1} for the above two transitions 
are, therefore, corroborate well with our assignments. Similarly, the multipolarities and the types of 
the transitions reported in Ref.\cite{a1} for all the in-band transitions in bands B1 and B2 corroborate 
well with the assignments based on the DCO and PDCO measurements in the present work. The spins and 
parities of most of the levels in $^{197}$Tl reported in \cite{a1}, are therefore, confirmed in the 
present work.

The spin and parity of the 1553 keV level and the 2114 keV band-head have been determined from the multipolarity 
and the type of the 557.5- and 561-keV transitions, respectively. These transitions were reported as mixed 
($M1+E2$) type and $E1$ type of transitions, respectively, by R.M. Lieder et al. in Ref.\cite{a1}. The 
perpendicular and the parallel spectra projected from the PDCO matrix for these two transitions, gated by 
561- and 557.5-keV $\gamma$ rays are shown in Figs.~6(a) and 6(b), respectively. It can be seen from these 
figures that the perpendicular counts (N$_\perp$) are clearly more than the parallel counts (N$_\parallel$) 
for both the transitions and hence, the value of $\Delta_{PDCO}$ are positive for both the transitions. This 
suggests that both the 557.5- and the 561-keV $\gamma$ rays are of electric type. It may be noted that 
because of the 561-keV gate in Fig.~6(a), there is no contamination of the 557.1 -keV transition from the 
21/2$^-$ level at 2597 keV which is of $E2$ type. Similarly, in Fig.~6(b), even if the gating transition
contains some contribution from 557.1-keV peak, the counts in the 561-keV peak will not be affected as the
561- and the 557.1 keV transitions are not in coincidence. 

The $R_{DCO}$ value of 1.03(10) obtained for the 
557.5-keV transition gated by the 561-keV known $E1$ transition (see Table~I) indicates that it is of same 
multipolarity as 561-keV, i.e dipole. We have also obtained R$_{DCO} = 1.70(40)$ for the 557.5-keV transition 
gated by the 767-keV stretched $E2$ transition which agrees well with the calculated value of $1.64$ assuming 
557.5-keV transition as a dipole transition. A value of R$_{DCO} = 1.76(35)$ has been obtained for the 561-keV 
transition using the 767-keV stretched $E2$ gate, confirming its dipole assignment. It may also be noted again 
that there is no contamination from the 557.1-keV transition in the 767-keV gate. Therefore, these measurements 
indicate that the type and the multipolarity of the 557.5-keV transition is same as that of the 561-keV 
transition. 

We have also calculated the value of polarization and PDCO using the formalism given in Ref.\cite{rp} (and 
references there in) for the 557.5-keV transition assuming it to be $E1$ and mixed $M1+E2$ transition with 
the mixing ratio as given in Ref.\cite{a1}. The measured value of +0.13(2) is in better agreement with the 
calculated value of +0.10 obtained assuming $E1$ transition rather than the calculated value of +0.04 
obtained assuming mixed $M1+E2$ transition for the 557.5-keV $\gamma$. The calculations also yields a value 
of +0.08 for the known $E1$ transition of 561-keV, for which the measaured value is +0.09(1). 

All the above results indicate that the 557.5-keV $\gamma$-ray is of $E1$ type rather than a mixed $M1+E2$ type 
and hence, the spin and parity of the 1553 keV level is 13/2$^+$. Also, as the 561-keV transition is an $E1$ 
transition (our results is in agreement with Lieder et al.), the spin and parity of the 2114 keV level is 
15/2$^-$. It may be noted that the angular distribution as well as the conversion electron measurements of the 
557.5-keV $\gamma$ ray from the 1553 keV level by Lieder et al. were not free from the contamination due to the 
557.1-keV $E2$ transition but there was no such contamination for the 561-keV transition from the 2114 keV 
level. The multipolarity of the 811-keV $\gamma$ ray, from the 2114-keV level to the 1303-keV level, has been 
found to be of mixed dipole ($M1+E2$) character from our DCO and the polarization measurements (see Fig.~3 
and Table~I). This lends additional support to the new $J^\pi$ assignments of the 2114- and 1553-keV levels 
in this work.

Better statistics for the higher lying $\gamma$ rays, obtained in the present work, helped us to establish 
definite $J^\pi$ assignments of the 3169- and 3277-keV states which were only tentatively assigned in the 
previous work \cite{a1}.

\begin{figure}[!]
\begin{center}
\includegraphics*[scale=0.3, angle = 0]{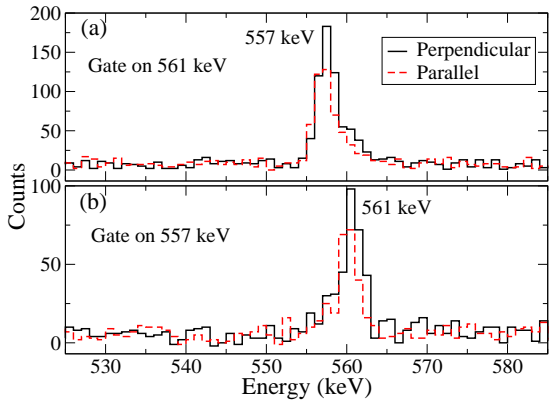}
\caption{(Color online) The perpendicular and the parallel spectra for the 557- and 561-keV lines gated by 
(a) 561- and (b) 557-keV $\gamma$ rays, respectively. It is clearly observed that the perpendicular counts 
are more than the parallel counts for both the $\gamma$ lines which indicate that both the transitions are 
of electric type.}
\label{Fig6}
\end{center}
\end{figure}

In one of the previous experiments ~\cite{a1a}, the $J^{\pi}$ of the 1953- and 1867-keV states were assigned 
as 9/2$^-$ and 9/2, respectively. In the present work the 958-keV $\gamma$ ray from the 1953-keV state to the 
995-keV state has been found to be of $E1$ type from the DCO and the polarization measurements. The 
polarization measurement for the weak 871-keV $\gamma$ ray from the 1867-keV level to the 995-keV level could not 
be measured, but the DCO ratio value for this $\gamma$ indicate that it is of mixed dipole character. As a 
result, the $J^{\pi}$ of the 1953- and 1867-keV levels have been assigned as 13/2$^+$ and 13/2$^-$, 
respectively. 

The known $\gamma$-rays in band B2 has been confirmed in the present work and the J$^\pi$ assignments have
been made. The tentatively-assigned 348- and 535-keV $\gamma$-rays in this band have been placed firmly in
this work from the coincidence relation with the other $\gamma$ rays. 
Neither the DCO ratio nor the PDCO ratio could be obtained for the weak cross-over transitions of energy
470 keV in band B1 and 348, 535, 614 keV in band B2. However, the spin and parity of the states, from which 
these transitions are emitted, are fixed from the mixed ($M1+E2$) transitions corresponding to those 
cross-over transitions. Therefore, $E2$ multipolarity has been adopted for the above transitions.

A new band, B3, has been observed for the first time in the present work. The $\gamma$ rays belonging to this 
band has been observed in the gated spectra shown in Fig.~4. These $\gamma$ rays are found to be of mixed 
$M1+E2$ in nature. The cross-over $E2$ transitions could not be assigned with certainty in this work as they 
overlap with the other known strong transitions. They can only be placed unambiguously from a double gated 
spectrum, obtained from a tripple $\gamma$ cube, which is outside the purview of the present work. The spin
and parity of the band-head of this band is based on the R$_{DCO}$ measurement of the 262-keV transition
which decays from the 2376-keV band-head to the 2114 keV level. To avoid contamination due to the 263-keV
and the 264-keV transitions from the 2529 keV and the 2860 keV levels, different gating transitions were
chosen to obtain $R_{DCO}$ values of these transitions. The $R_{DCO}$ value of the 262-keV transition indicate 
that it is a $\Delta I = 1$ transition and hence the band-head spin of 17/2 is assigned for the band B3. However,
as the 262-keV transition was assigned as an $M1+E2$ transition, only tentatively, the parity of this band is 
also tentatively assigned as negative. 

\section{Discussion}

\begin{figure}
\begin{center}
\includegraphics*[scale=0.3, angle = 0]{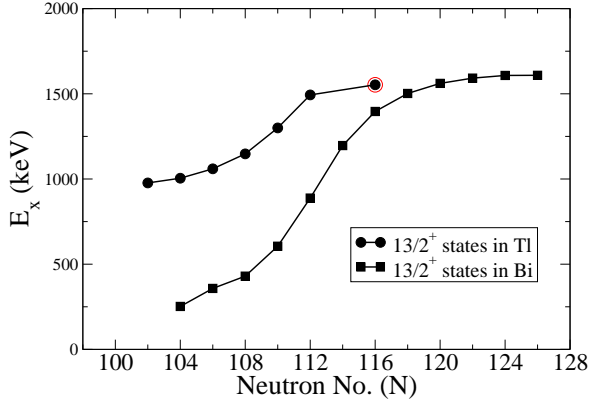}
\caption{(Color online) Excitation energy ($E_{x}$) of the $13/2^{+}$ states in Tl and Bi isotopes as a 
function of neutron number. The (red) encircled one corresponds to that in $^{197}$Tl from the present work.}
\label{Fig7}
\end{center}
\end{figure}

The proton Fermi level for the Tl isotopes lie below the $Z=82$ shell closure near the $2s_{1/2}$ 
and the $3d_{3/2}$ orbitals whereas, the  neutron Fermi level lies near the $i_{13/2}$, $f_{5/2}$ and 
$p_{3/2}$ orbitals. Therefore, the 1/2$^+$ ground state and the 3/2$^+$ first excited state in
the $^{197}$Tl are due to the occupation of the odd-proton in the $s_{1/2}$ and $d_{3/2}$ orbitals. 
The 9/2$^-$ isomeric states, observed in the neutron deficient odd-mass Tl isotopes in the mass region $A=190$, 
have been interpreted in terms of the occupation of the odd proton in the $h_{9/2}$ ``intruder" 
orbital. The [505]9/2$^-$ Nilsson orbital comes down in energy for oblate deformation and becomes available 
for the odd proton in Tl nuclei. Consequently, the observed band structures based on the 9/2$^-$ isomeric states 
in the odd-$A$ Tl nuclei, and similar to the band B1 in $^{197}$Tl, were interpreted as oblate deformed band. 
Similarly, the $\pi$[606]$13/2^{+}$ orbital originated from the $\pi i_{13/2}$ state also intrudes in to 
the Fermi level for Tl nuclei for such deformation. This is evident from the observation of the 13/2$^+$
state in the odd-$A$ neutron deficient Tl isotopes. Band structure based on this state has been reported
for the lighter isotopes of $^{191,193}$Tl \cite{a4,a5}. However, this state has not been reported so far 
in $^{195,197}$Tl. 

In the present work, we have observed two 13/2$^+$ states at 1553 keV and 1953 keV; either of which may be 
a candidate for the $\pi i_{13/2}$ state. The decay pattern of the former 13/2$^+$ state in $^{197}$Tl is 
quite similar (a strong $E1$ transition to the h$_{9/2}$ band) to the observed decay pattern of the 
$\pi i_{13/2}$ states in the lighter Tl isotopes \cite{183Tl,185Tl,a3,new74,a5}. Therefore, the 13/2$^+$ state 
at 1553 keV in $^{197}$Tl seems to be a candidate for the $\pi$[606]$13/2^{+}$ ``intruder" configuration. While 
collective structure based on this configuration have been observed for the lighter Tl iostopes, single-particle 
interpretation was given for the band-like sequence based on the 13/2$^+$ level in $^{193}$Tl \cite{a5}. No 
collective band structure based on this state, however, has been observed in the present work for $^{197}$Tl.

The variations of the excitation energies of the lowest $13/2^+$ states (most likely originated from the
$\pi$[606]$13/2^{+}$ ``intruder" configuration) in the odd-$A$ Tl isotopes are shown in Fig. 7 as a function 
of the neutron number. The same for the Bi isotopes are also shown in this figure for comparison. The 
excitation energy of the 13/2$^+$ state in $^{197}$Tl, shown as a red encircled point, follows the 
systematic behaviour of the odd-$A$ Tl isotopes similar to the case for Bi isotopes. For the lighter isotopes 
of Tl, the excitation energies of the 13/2$^+$ states are at much higher energies compared to those in Bi 
isotopes. This is because of the fact that the proton Fermi level of Bi isotopes lies near the $h_{9/2}$ 
orbital above the $Z=82$ shell closure and so, for relatively larger oblate deformation (for the lighter Bi 
isotopes) the energy difference between the $\pi h_{9/2}$ and the $\pi i_{13/2}$ states is smaller compared 
to that at the smaller deformation (for the heavier Bi isotopes). However, the energy of this state relative 
to the $\pi h_{9/2}$ state should be similar for both Tl and Bi nuclei, which is indeed true for the $N = 116$ 
isotones of Bi and Tl. As the neutron number increases, the excitation energy of the 13/2$^+$ state is 
observed to increase for both Bi and Tl and tends to saturate, suggesting near spherical shapes for the heavier 
isotopes.

\begin{figure}[!]
\begin{center}
\includegraphics*[scale=0.3, angle = 0]{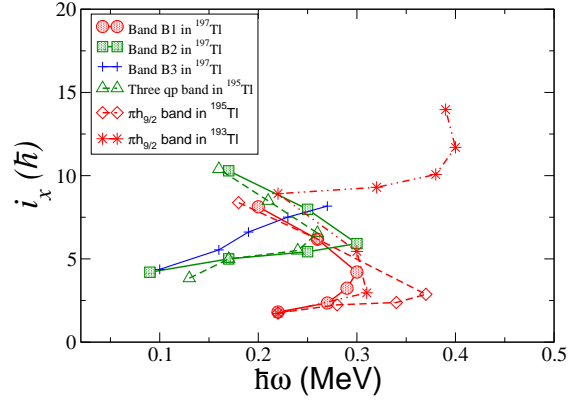}
\caption{(Color online) Experimental alignments ($i_x$) as a function of the rotational frequency 
($\hbar\omega$) for the $\pi h_{9/2}$ band in $^{193}$Tl~\cite{a5}, $^{195}$Tl~\cite{a1} and $^{197}$Tl 
along with that for the three quasiparticle (qp) band in $^{195}$Tl~\cite{a1} and $^{197}$Tl. The Harris 
reference parameters are chosen to be $J_{0}$ = 8.0{$\hbar$}$^{2}$ $MeV^{-1}$ and $J_{1}$ = 40
{$\hbar$}$^{4}$ $MeV^{-3}$.}
\label{Fig8}
\end{center}
\end{figure}

The alignments ($i_x$) for the bands in $^{197}$Tl, observed in the present work has been shown in Fig. 8 
as a function of rotational frequency $\hbar\omega$ along with those in the neighboring isotopes $^{193}$Tl 
and $^{195}$Tl. It can be seen from this figure that the initial alignments for the $\pi h_{9/2}$ band of 
all the three isotopes are very similar. The experimental band crossing of this band in $^{197}$Tl, 
$^{195}$Tl and $^{193}$Tl~\cite{a5} take place at the rotational frequencies of $\hbar\omega \sim 0.30$ MeV, 
0.36 MeV and 0.28 MeV, respectively. Although, there are differences in the crossing frequencies, the gain 
in alignments after the band crossing are seen to be very similar for the three isotopes. The first band 
crossing observed in $^{193}$Tl~\cite{a5} was interpreted as due to the neutron pair alignments in the 
$i_{13/2}$ orbital. In comparison to that, the observed band crossing of band B1 in $^{197}$Tl may also be 
attributed to the neutron pair alignment in $\nu$$i_{13/2}$ orbital. Therefore, the configuration of the 
higher lying states in this band (above 2426 keV) would be $\pi h_{9/2} \otimes \nu i_{13/2}^{2}$.

The excitation energies of the bands B2 and B3 in $^{197}$Tl indicate that these negative parity bands are 
based on the three quasiparticle (qp) configurations. The similarities in the values of $i_x$ of these bands 
indicate that they are of similar configuration. In the neighboring even-even Hg nuclei, the negative parity 
bands have been interpreted as two-qp decoupled bands with a microscopic structure of a completely decoupled 
$i_{13/2}$ neutron and a low-j neutron of opposite parity in the adjacent $p_{3/2}$, $p_{1/2}$ or $f_{5/2}$ 
orbitals \cite{hg1,hg2,hg3}. 

The alignment pattern for the band B2 in $^{197}$Tl is quite similar to that for the 5$^-$ band in its 
immediate even-even neighbor $^{196}$Hg. The alignment of this band in $^{196}$Hg was interpreted as the 
coupling of a fully aligned $i_{13/2}$ neutron and a poorly aligned low-j ($p_{3/2}$, $f_{5/2}$ or 
$p_{1/2}$) neutron \cite{hg3}. Further alignment of two more $i_{13/2}$ neutrons has also been observed 
for this band in $^{196}$Hg at a higher spin with a gain in alignment of about 9$\hbar$. The similarities in 
the values of $i_x$ between the 5$^-$ band in $^{196}$Hg (before band crossing) and the bands B2 and B3 in 
$^{197}$Tl indicate that their intrinsic structures are similar i.e one neutron in the $i_{13/2}$ orbital and 
another in a low-j negative parity orbitals. Since most of the observed aligned angular momenta are taken care 
by these neutron configurations, the contribution of the odd-proton to the $i_x$ is very small and hence the
odd-proton should be in a high-$\Omega$ orbital of $h_{9/2}$ or $i_{13/2}$, which is indeed the case for
oblate deformation. Considering the negative parity of the bands, the intrinsic configuration of the bands 
B2 and B3 in $^{197}$Tl may be considered as $\pi i_{13/2} \otimes \nu i_{13/2} \otimes (p_{3/2}, p_{1/2}, 
f_{5/2})$. A strong overlap of these bands to the $\pi i_{13/2}$ state through an $E1$ transition supports 
this assignment. It may be noted that for either case of the odd-proton be in $h_{9/2}$ or $i_{13/2}$ orbital, 
the contribution of the odd-proton to the aligned component of the angular momenta would be very small as it is 
in the high-$\Omega$ orbital. So, the plot of $i_x$ would look similar for both the cases and it would not be 
possible to distinguish the two different configurations from the aligned angular momenta. The three-qp level 
structures above the 13/2$^+$ state in $^{193}$Tl are somewhat different than those for the $^{195,197}$Tl. In 
$^{193}$Tl, several band structures or sequence of levels have been observed to decay to the 17/2$^+$ state 
which is not the case for $^{195,197}$Tl. It seems that the configuration of the three-qp band in $^{197}$Tl 
is somewhat different from that in $^{193}$Tl although the aligned angular momenta ($i_x$) are similar 
in these isotopes. 
However, the three-qp band in $^{195}$Tl has been reported to be of positive parity \cite{a1} in contrast to 
the negative parity in $^{197}$Tl. Therefore, it may be worthwhile to confirm the parity assignment of the 
three-qp band in $^{195}$Tl in a separate experiment.  

The experimental band crossings of the band B2 have been observed to take place at $\hbar\omega \sim 0.30$ 
MeV which is little higher than that in $^{195}$Tl (at $\hbar\omega\sim 0.26$ MeV). No band crossing has 
been observed for the band B3 which may indicate that the crossing is further delayed in this band. However, 
a gradual gain in alignment may be noticed for the band B3. It would be interesting to extend this band to 
higher spin members to see if the gradual alignment continues or a band crossing occurs at a higher 
rotational frequency.

An alternative interpretation of the band B2 and B3 may be given based on the recent results obtained for
the excited states in the immediate even-even neighbor $^{196}$Hg. The excited states in this nucleus have 
been studied by ($\alpha$,x$n$) reaction up to 2.4 MeV of excitation energy and the states (up to 1.9 MeV) 
have been interpreted using a theoretical description within the $U_\nu (6/12)\otimes U_\pi (6/4)$ extended 
supersymmetry model \cite{196hg}. In this model, $^{196}$Hg has been considered as a fifth supermultiplet 
member to the so-called magical quartet consisting of $^{194,195}$Pt and $^{195,196}$Au. In their study, 
a low-lying 3$^+$ state and a second 4$^+$ state have been identified in $^{196}$Hg nucleus through precise 
spin determination. The band-head spin-parity of 15/2$^-$ and 17/2$^-$ for the bands B2 and B3, respectively, 
in $^{197}$Tl may be considered as the $h_{9/2}$ proton coupled to the low-lying 3$^+$ and 4$^+$ states of 
its immediate even-even neighbour $^{196}$Hg. The difference in the excitation energies between the 4$^+$ 
and the 3$^+$ states in $^{196}$Hg ($\sim 300$ keV) is found to be very similar to that between the band 
heads of the bands B3 and B2 (262 keV) in $^{197}$Tl. Therefore, the configuration of the bands B2 and B3 
may be considered as 3$^+_{^{196}\text Hg} \otimes \pi h_{9/2}$ and 4$^+_{^{196}\text Hg} \otimes \pi 
h_{9/2}$, respectively. It would be interesting to investigate if the four neighboring nuclei $^{196,197}$Hg 
and $^{197,198}$Tl can be described as the members of a supermultiplet.

\subsection{TRS calculations}
\begin{figure}[!]
\begin{center}
\includegraphics*[scale=0.345, angle = 0]{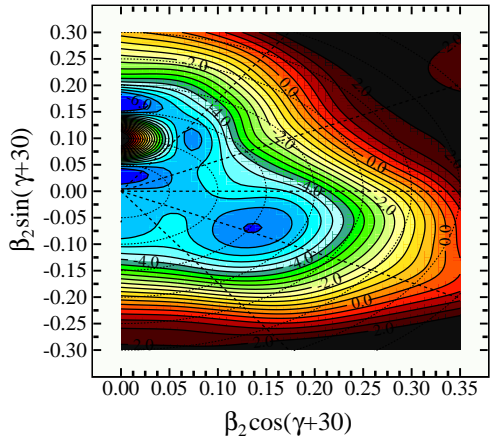}
\caption{(Color online) Contour plots of the total Routhian surfaces (TRSs) in the $\beta_2$ - $\gamma$ 
deformation mesh for the $\pi h_{9/2}$ configuration corresponding to the band B1 in $^{197}$Tl. The 
contours are 400 keV apart.}
\label{Fig9}
\end{center}
\end{figure}

\begin{figure}[!]
\begin{center}
\includegraphics*[scale=0.345, angle = 0]{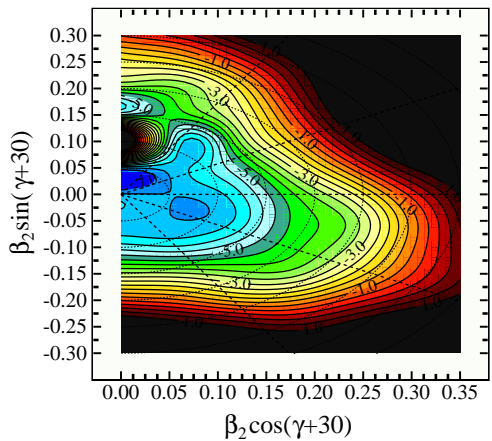}
\caption{(Color online) Same as Fig.7 but for the $\pi i_{13/2}$ configuration corresponding to the 
$13/2^{+}$ state in $^{197}$Tl.}
\label{Fig10}
\end{center}
\end{figure}

The total Routhian surface (TRS) calculations have been performed for the $\pi h_{9/2}$ (band B1) and 
the $\pi i_{13/2}$ configurations in $^{197}$Tl. The Hartee-Fock-Bogoliubov code of Nazarewicz~{\it 
et al.} \cite{anaza1,anaza2} was used for the calculations. The procedure has been outlined in 
Ref.~\cite{agm2,hp,hp1}. A deformed Woods-Saxon potential and pairing interaction was used with the 
Strutinsky shell corrections method. The TRSs were calculated in the $\beta_2 - \gamma$ deformation 
mesh points and minimized in the hexadcapole deformation $\beta_4$. The contour plots of the TRSs, 
calculated for the band B1 have been shown in Fig. 9. In these plots, $\gamma = 0^o$ corresponds to a
prolate shape and $\gamma = -60^o$ to an oblate shape. The value of $\gamma$ lies in between these two 
limits for a triaxial shape. Fig. 9 clearly shows a minimum in the TRS at a deformation of $\beta_2 
\sim 0.15$ and $\gamma \sim -58^o$, indicating an oblate shape for $^{197}$Tl for the $\pi h_{9/2}$ 
configuration. Therefore, the rotational band observed in $^{197}$Tl based on this configuration is 
due to the oblate deformation as in the other lighter Tl isotopes. The surfaces calculated for the 
$\pi i_{13/2}$ configuration in $^{197}$Tl have been shown in Fig. 10. The minimum in the Routhian 
surfaces, in this case, has been found to be at $\beta_2 \sim 0.08$ and $\gamma \sim -50^o$, indicating 
an oblate shape with very small deformation. Rotational band like structure is not expected for such a
small deformation. It may be noted that no rotational band structure has been observed in $^{197}$Tl 
based on the $\pi i_{13/2}$ state in accordance with the TRS calculations. 

\section{\bf Summary}
The $\gamma$-ray spectroscopy of the odd-$A$ $^{197}$Tl has been studied in the fusion-evaporation reaction 
of $^{197}$Au target with $^{4}$He beam at 48 MeV. A new and improved level scheme of $^{197}$Tl is 
presented from the present work which includes 8 new $\gamma$-ray transitions. The previously known 
$\gamma$-rays have been observed, some of the tentatively placed $\gamma$-rays have been confirmed and 
a new band structure has been identified in this nucleus. The DCO ratio and the polarization asymmetry 
ratio measurements have been carried out to assign the spins and parities of the levels. Evidence for the
existence of the $\pi i_{13/2}$ intruder state has been observed for the first time in this nucleus. However, 
no rotational band structure has been observed based on this state. A negative parity for the band B2 has been 
assigned in this work from the polarization measurements and a new configuration has been assigned for this 
three-quasiparticle band. The new band B3 has been found to be based on 17/2$^{(-)}$ band-head. Possible 
configurations of the bands have been discussed in the light of the neighboring nuclei. Configuration dependent 
TRS calculations have been performed for $^{197}$Tl. The observation of rotational band based on $\pi h_{9/2}$ 
configuration and the non-observation of any rotational band based on $\pi i_{13/2}$ configuration are well 
corroborated by the deformation parameters obtained from the TRS calculations.

\section{\bf Acknowledgement}
The untiring effort of the cyclotron operators at VECC to provide a good $\alpha$-beam is gratefully 
acknowledged. We would also like to thank UGC-DAE-CSR-KC for lending one HPGe detector to us.
Fruitful discussion with Prof. Robert V.F. Janssens is greatefully acknowledged.\\

\end{document}